\documentclass[sigconf,edbt]{acmart-edbt2025}
\def\BibTeX{{\rm B\kern-.05em{\sc i\kern-.025em b}\kern-.08em
    T\kern-.1667em\lower.7ex\hbox{E}\kern-.125emX}}

\usepackage{booktabs} 

\setcopyright{rightsretained}

\acmDOI{}

\acmISBN{978-3-89318-099-8}

\acmConference[EDBT 2025]{28th International Conference on Extending Database Technology (EDBT)}{25th March-28th March, 2025}{Barcelona, Spain}
\acmYear{2025}

\usepackage{amsmath}
\usepackage{mathtools}
\usepackage{graphicx} 
\usepackage{url}
\usepackage{enumitem} 

\usepackage{stmaryrd} 

\usepackage{upgreek} 
\usepackage{bbding} 

\usepackage{listings} 
\usepackage{tabularx} 
\usepackage{booktabs} 
\usepackage{multirow}
\usepackage{hhline} 
\usepackage{diagbox}
\usepackage{pgfplotstable} 

\usepackage{xcolor,colortbl}    

\usepackage{tikz} 
\usetikzlibrary{matrix}
\usetikzlibrary{calc}
\usetikzlibrary{math}
\usetikzlibrary{arrows.meta,positioning}
\usetikzlibrary{intersections, pgfplots.fillbetween}

\usepackage{easybmat} 

\usepackage{adjustbox}
\def\eox{\unskip\kern 10pt{\unitlength1pt\linethickness{.4pt}$\diamondsuit${}}} 

\usepackage{rotating}		

\usepackage{xspace} 

\usepackage{subcaption} 
\newcommand{\hide}[1]{}

\usepackage[ruled,noend,linesnumbered]{algorithm2e} 

\usepackage{setspace} 

\DontPrintSemicolon     
\SetNlSty{}{}{}                
\SetAlgoInsideSkip{smallskip}   

\SetAlFnt{\small}			
\SetAlCapFnt{\small}		
\SetAlCapNameFnt{\small}

\hypersetup{                                                            
	pdfpagemode=UseNone,               
    pdfstartview=Fit,
    pdfpagelayout=SinglePage,       
    bookmarks=false,
    bookmarksnumbered=true,
    bookmarksopen=false,             
    pdftex=true,
    unicode,
    colorlinks=true,       
    linkcolor=blue,          
    citecolor=cyan,  
    filecolor=magenta,      
     urlcolor=blue           
}

\usepackage{aliascnt}  	

\newtheorem{questionW}{Question}
\newtheorem{resultW}{Result}

{\end{itshape}
}
\AtEndPreamble{%
    \hypersetup{colorlinks,
      linkcolor=purple,
      citecolor=blue,
      urlcolor=ACMDarkBlue,
      filecolor=ACMDarkBlue}}

\usepackage{tcolorbox}		
\tcbuselibrary{breakable,skins}		

\tcbset{examplestyle/.style={
		enhanced jigsaw,	
		colback=blue!08,	
		colframe=blue!08,	
		arc=2mm,
		boxrule=0pt,		
		left=1mm,
		right=1mm,
		left skip= 0mm,  
		right skip= 0mm, 
		top=1mm,		
		bottom=1mm,		
		breakable,		
		parbox = false,		
		before={\par\pagebreak[0]\vspace{1mm}\parindent=0pt},		
		after={\par\pagebreak[0]\vspace{1mm}\parindent=0pt},				
		bottomrule = 0mm,
		boxsep = 0mm,					
		topsep at break=0pt,			
		bottomsep at break=0pt,			
		pad at break=0mm,
		pad before break=1mm,		
		pad after break=1mm,		
		bottomrule at break=0mm,
		toprule at break=0mm,		
		}}

\usepackage{footnote}

\tcolorboxenvironment{example}{examplestyle}

\makeatletter
\DeclareRobustCommand*\uell{\mathpalette\@uell\relax}
\newcommand*\@uell[2]{
  \setbox0=\hbox{$#1\ell$}
  \setbox1=\hbox{\rotatebox{10}{$#1\ell$}}
  \dimen0=\wd0 \advance\dimen0 by -\wd1 \divide\dimen0 by 2
  \mathord{\lower 0.1ex \hbox{\kern\dimen0\unhbox1\kern\dimen0}}
}
\makeatother

\usepackage{marginnote}

\newcommand{\introparagraph}[1]{\textbf{#1.}} 

\renewcommand{\epsilon}{\varepsilon} 

\usepackage{scalerel}
\usepackage{tikz}
\usetikzlibrary{svg.path}

\definecolor{orcidlogocol}{HTML}{A6CE39}
\tikzset{
  orcidlogo/.pic={
    \fill[orcidlogocol] svg{M256,128c0,70.7-57.3,128-128,128C57.3,256,0,198.7,0,128C0,57.3,57.3,0,128,0C198.7,0,256,57.3,256,128z};
    \fill[white] svg{M86.3,186.2H70.9V79.1h15.4v48.4V186.2z}
                 svg{M108.9,79.1h41.6c39.6,0,57,28.3,57,53.6c0,27.5-21.5,53.6-56.8,53.6h-41.8V79.1z M124.3,172.4h24.5c34.9,0,42.9-26.5,42.9-39.7c0-21.5-13.7-39.7-43.7-39.7h-23.7V172.4z}
                 svg{M88.7,56.8c0,5.5-4.5,10.1-10.1,10.1c-5.6,0-10.1-4.6-10.1-10.1c0-5.6,4.5-10.1,10.1-10.1C84.2,46.7,88.7,51.3,88.7,56.8z};
  }
}

\RequirePackage{etoolbox}
\DeclareRobustCommand\orcidicon[1]{\href{https://orcid.org/#1}{\mbox{\scalerel*{
\begin{tikzpicture}[yscale=-1,transform shape]
\pic{orcidlogo};
\end{tikzpicture}
}{|}}}}

\usepackage{array}
\usepackage{makecell}
\usepackage[utf8]{inputenc}

\begin{document}
\title{Model Lakes}

\author{Koyena Pal}
\affiliation{%
  \institution{Northeastern University}
  \city{Boston}
  \state{Massachusetts, USA}
}
\email{pal.k@northeastern.edu}


\author{David Bau}
\affiliation{%
  \institution{Northeastern University}
  \city{Boston}
  \state{Massachusetts, USA}
  }
\email{davidbau@northeastern.edu}

\author{Ren\'ee J. Miller}
\affiliation{%
  \institution{Northeastern \& U. Waterloo}
  \city{Waterloo}
  \state{ON, Canada}
  }
\email{rjmiller@uwaterloo.ca}

\renewcommand{\shortauthors}{}

\begin{abstract}
Given a set of deep learning models, it can be hard to find models appropriate to a task, understand the models, and characterize how models are different one from another. Currently, practitioners  rely on manually-written documentation to understand and choose models.
However, not all models have complete and reliable documentation. As the number of models increases, the challenges of finding, differentiating, and understanding models become increasingly crucial. Inspired from research on data lakes, we introduce the concept of {\em model lakes}.  We formalize key model lake tasks, including {\em model attribution, versioning, search, and benchmarking}, and discuss fundamental research challenges in the management of large models. We also explore what data management techniques can be brought to bear on the study of large model management.
\end{abstract}

%
%



\maketitle
\section{Introduction}
With the dramatic rise in AI capabilities across a variety of domains~\cite{deng2023mindweb, schick2024toolformer, openaichat,  bardchat,elsaai, seeingai, fitnessai, lensa, socratic, trummer2022codexdb, chintagunta2021medically}, many organizations have begun to commit significant resources to developing Machine Learning Models.  Many of these are fine-tuned versions of popular foundation models, such as Llama-3~\cite{DBLP:journals/corr/abs-2406-07536}, Mistral~\cite{DBLP:journals/corr/abs-2310-06825}, DeepSeek-R1~\cite{deepseekai2025deepseekr1incentivizingreasoningcapability}, 
Stable Diffusion~\cite{DBLP:conf/cvpr/RombachBLEO22}, BART~\cite{lewis-etal-2020-bart}, and BERT~\cite{devlin-etal-2019-bert}. Proprietary closed-source models such as GPT-4~\cite{gpt4}, Gemini~\cite{gemini} and Claude-3~\cite{claude3} also support creation of fine-tuned models. To support this proliferation, sharing, and reuse of large models, many models are hosted on platforms to support the collaborative use and sharing of models such as Hugging Face~\cite{hugging} and Kaggle~\cite{kaggle}.

As the number of pre-trained models grows, comparing them and selecting the right one for specific tasks becomes increasingly difficult (see Example~\ref{ex:model_search_example}). Documentation, particularly model cards~\cite{modelcardpaper}, aims to provide essential insights, but ~\citet{liang2024whats} have revealed a concerning lack of transparency and completeness in these resources. This makes it hard for users to make informed decisions, especially when navigating model sharing platforms. Efforts such as the Data Provenance Initiative~\cite{dataprovenanceinitative}, MLCommons~\cite{mlperfinferencebenchmark, mlperftrainingbenchmark}, and  Responsible Foundation Model Development Cheatsheet~\cite{modeldevcheatsheet} have been introduced to enhance the documentation of model creation and capabilities, with a particular focus on detailing their training data. However, not all model creators adhere to these guidelines, meaning that many existing and future models may still lack crucial information needed by users. To address this gap, we propose a systematic breakdown and formalization of tasks for ``model lakes" --- a system containing numerous heterogeneous pre-trained models and related data in their natural formats (one example is Hugging Face~\cite{hugging}).

We introduce {\em model lakes}, as a parallel to data lakes, and discuss how important innovations in data lakes~\cite{DBLP:journals/pvldb/NargesianZMPA19}, including data discovery, annotation, and version management, can (and should) be applied within model lakes and studied with the same vigor. We look at how model lakes are currently managed and define and discuss tasks that can be used to better inform users about the models and their relationships.  Others have discussed how one type of model (LLMs) may disrupt data management~\cite{fernandez2023visionpaper}. We focus on how data management can transform the management and use of AI models.
\begin{example}
\label{ex:model_search_example}
    \textbf{Model Search Problem:} Consider a situation where a user wants to find a model that can summarize a legal document and simplify it in a non-technical manner. On Hugging Face (as of Sept 2024), the user finds that there are around 1M+ models uploaded and 1950 of them have the `summarization' task tag. While there are filters (trending, most likes, most downloads, model name search, and more), the user finds it hard to choose which model to use. There are various concerns that the user goes through as she scrolls through different model cards (a common semi-structured form of model documentation).  Is this model aware of legal jargon? Is it good at summarizing and simplifying legal documents? Is this the latest version of the model? Was this model trained on legal texts and if so which texts?  What are other models that are similar to this model? Are they also trained on the same or different legal texts?
\end{example}

\section{Three Viewpoints of a Model}\label{sec:model-viewpoints}

An AI model, $\mathcal{M}$, can be analyzed from three viewpoints: according to its \textbf{history}, \textbf{intrinsic} composition, or \textbf{extrinsic} behavior. These viewpoints highlight different aspects of the model’s characteristics, aspect that we will show are useful in analyzing model lake problems and solutions.

The \textbf{history} of the model is defined by its training data ($\mathcal{D}$) and training algorithm ($\mathcal{A}$), which may include processes like fine-tuning, model editing, or other adaptation techniques. The degree to which history is documented within a model lake can vary greatly~\cite{liang2024whats}.

The \textbf{intrinsic viewpoint} concerns the model's internal structure. This includes the model architecture ($f_{*}$), and the specific trained parameters ($\theta$). The architecture refers to the structural design of the model, such as a combination of multi-layer perceptrons (MLPs) and attention layers in transformer models.  Formally, the architecture defines a function family $f_*$ which is instantiated with the specific learned parameters $\theta$ to create the function $f_\theta.$

In contrast, the \textbf{extrinsic viewpoint} focuses on the model’s observable behavior and performance in user-defined tasks. The model parameters, $\theta$, are not visible as part of the extrinsics. However, the function $f_\theta$ and the model behavior, $p_\theta$, is extrinsic. For example, in the case of an unconditional generative model, the extrinsics correspond to the observable probability distribution defined by the model, $p_{\theta}(x)$.  For a classifier, the extrinsics are defined by the behavior of the modeled predictions $p_{\theta}(y|x) =  f_{\theta}(x)$.  Extrinsics can be observed in terms of a neural network's action on inputs and outputs $x$ and $y$, without requiring any knowledge of its training data or its internal structure.

\begin{figure}[!ht]
    \centering
\includegraphics[width=\columnwidth]{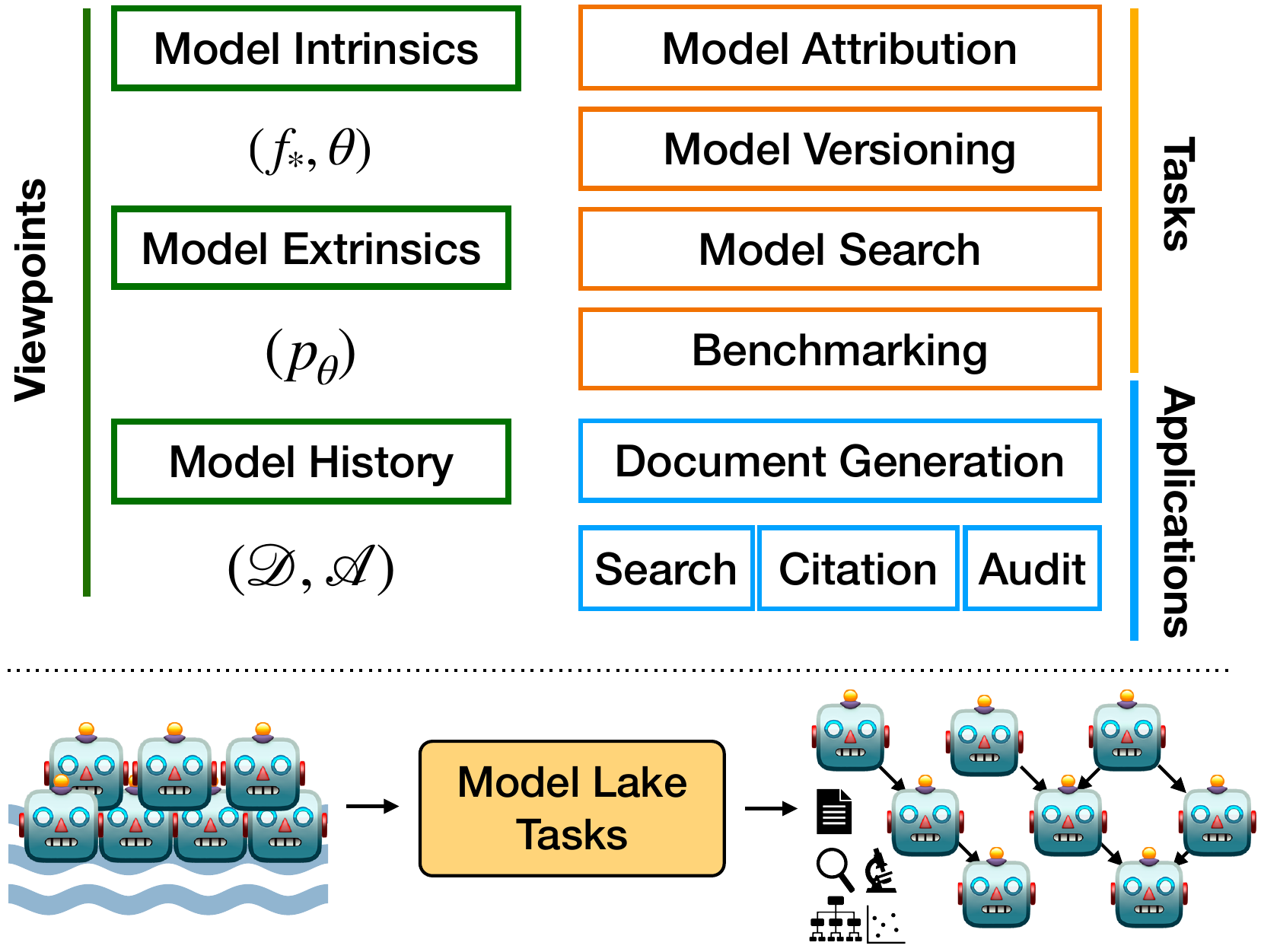}
    \caption{On the bottom of the figure, we illustrate the concept of \textit{model lakes}, where diverse models are stored.  
    As these models undergo the tasks outlined on the top-right side, users gain a deeper understanding of their origins, strengths, and how they are structured in relation to other models. This process provides key insights into the models' development, performance capabilities, and their positioning within the broader landscape of models. A model is defined as $\mathcal{M} = (\mathcal{D}, \mathcal{A}, f_*, \theta, p_\theta)$, where $\mathcal{D}$ (training data) and $\mathcal{A}$ (algorithm) can be traced through documentation, while architecture $f_*$ and parameters $\theta$ come from accessible model weights, and behavior $p_\theta$ from observable outputs (illustrated on the upper left side of the figure).}
    \label{fig:model-lakes-concept}
\end{figure}

The distinction among intrinsics, extrinsics, and history is useful because, while every model $\mathcal{M} = (\mathcal{D}, \mathcal{A}, f_*, \theta, p_\theta)$ has all these characteristics, there are cases where certain aspects may be unavailable. For example, in a Model Lake, the intrinsic details of some models might be inaccessible, and some analysis methods may rely solely on extrinsic observations or historical records to understand a model’s behavior. We use this distinction to analyze  possible solutions to a variety of model lake tasks.  Our envisioned model lake is depicted in~\Cref{fig:model-lakes-concept}.

\section{Formalizing Model Lake Tasks}\label{sec:model-lake-tasks}
Model lake tasks are specifically concerned with gathering and presenting insights about the models that are stored within the lake.  Topics related to the infrastructure behind model training or storage systems fall outside this scope. 

\smallskip
\noindent
\introparagraph{Model Attribution}
In data lakes, data provenance is the ``description of the origins of a piece of data and the process by which it arrived in a database"~\cite{buneman2001and}.  Model provenance (often called attribution) considers questions like ``Why was image X generated when the model was given input Y?"
If the history is recorded, the history can be consulted to directly ask the \emph{training data attribution} question formally: which training data items $d \in \mathcal{D}$ are most influential on the decision; in other words, which $d$, if they were not present in the training data, would cause the decision to change the most?
When history is not available, attribution can be studied using intrinsic and/or extrinsic clues to provide insight into the attribution of model decision behavior.  For example, we can perform \emph{sensitivity analysis} on the observable extrinsics of a model by asking: which aspects of the inputs to $f_\theta$ or $p_\theta$ are most important in a model's prediction of a particular output?  And if we have access to model intrinsics, we can study \emph{feature} or \emph{representation analysis}, which asks, which internal representations or internal ``concepts'' within the model are most important for a decision?

\smallskip
\noindent
\introparagraph{Model Versioning}
In the model versioning task, we want to understand whether (and possibly how) a model has been created from other models (similar to studying how a version of a data set or table may have been derived from another~\cite{DBLP:journals/pvldb/ShragaM23}). One possible definition of model versioning that uses an intrinsic viewpoint~\cite{horwitz2024origin} is:
Given a model, $\mathcal{M}_t$ and a set of $N$ models, $\{\mathcal{M}_{c} | c \in N\}$, construct a directed Model Graph, $\mathcal{T}$, 
where a directed edge between models indicates that one model is a version of the other.
The edges can describe the transformation. This can include training techniques, optimization techniques, as well as the data used to further train (fine-tune) models. An important problem in versioning is:  given a model's training parameters $\theta_t$ and a candidate model's  parameters, $\mathcal{\theta}_c$, is $\mathcal{\theta}_c$ a source of $\mathcal{\theta}_t$? Being a source model would mean that $\mathcal{\theta}_c$ or a version of $\mathcal{\theta}_c$ was used for training $\mathcal{\theta}_t$.

\smallskip
\noindent
\introparagraph{Model Search}
Model search is the task of finding a related or desired model.  Again we can consider this task from different viewpoints.  An extrinsic view considers the behavior of the model. Given a task function, $Q: X \rightarrow Y$ where a task takes an input, $x \in X$ and produces an output $y \in Y$, we want to find the best performing model, $\mathcal{M}_{best}$, for the given task among a set of $N$ candidate models, $U = {\mathcal{M}_1, \mathcal{M}_2, .... \mathcal{M}_N}$. Each candidate model is characterized by their observable behavior, i.e., $\{p_{1\theta},p_{2\theta}, ...,p_{N\theta} \}$. The optimal model is selected based on a scoring function of how close the model's behavior is to a query model.

Even considering only an extrinsic view, there are many formulations of model search.  If the goal is to find models that perform similarly on specific data  (for example, a specific image), then a possible definition is~\cite{Lu_2023}:  given a
point, $d$, and a set of $N$ candidate models, $U = {\mathcal{M}_1, \mathcal{M}_2, .... \mathcal{M}_N}$, rank models based on their similarity of their respective observable behavior $\{p_{1\theta},p_{2\theta}, ...,p_{N\theta} \}$ w.r.t. to $d$. This could be extended to other types of models where for a given model, $\mathcal{M}_t$ and a set of candidate models, $\{\mathcal{M}_{c}|c \in n\}$, we want to find \textbf{related models} based on a ranking function that considers the similarity of models' output distribution and semantic concepts/patterns present within them.

An intrinsic view of model search would find models with similar model architectures and training parameters. Or as done in data lakes where intrinisic search is the norm, we could create embeddings representing the important features of the model and design a fast nearest neighbor search over these embeddings. Considering model history, we could search for models that have been trained on the same dataset.  This is straight-forward when history is recorded, but when it is not fully explicit, we may leverage extrinsic or intrinsic clues in the search.

\smallskip
\noindent
\introparagraph{Benchmarking}
For single model tasks, a benchmark, $\mathcal{B}$ (for example, a set of images), is used to measure the performance of a model, $\mathcal{M}$ (or set of models) based on a scoring function, $S(\mathcal{M}, \mathcal{B}) \in R$.  For model lake tasks, we will need new (shared) model lake benchmarks (large sets of models $\mathcal{L}$) that can be used to measure the quality of a lake task solution.  This means that within a benchmark lake, we will need verified ground truth.  
\section{State-of-the-art in Model Mgmt}\label{sec:sota}
\introparagraph{Model Repositories, Registries, and Lakes} 
A model repository is a storage system for machine learning models. A model registry goes beyond basic storage by offering version control (the representation of versions, not today the discovery of version relationships mentioned in Section~\ref{sec:sota}). These registries typically enforce clear naming conventions (for models and versions) and organize structured, standardized metadata~\cite{neptuneai, sagemaker}. Recent advancements have added functionalities like explainable network intrusion detection~\cite{10733700} for security and integrations of different tools such as model monitoring and experiment tracking into a single framework~\cite{10425558}. Model registries are typically private, facilitating collaboration within organizations, while open platforms such as Civitai~\cite{civitai}, Hugging Face~\cite{hugging}, and Kaggle~\cite{kaggle} enable public sharing of models. These platforms, which we can consider as model lakes, assist users in model exploration by providing curated catalogs and keyword search, often leveraging both manually created model names and metadata for more efficient discovery. Unlike lifecycle creation and training platforms, model lakes focus on managing a set of AI models, including their interrelationships. This is distinct from model management in databases, which deals with schemas and their mappings~\cite{10.1145/1247480.1247482}. Current model lakes capabilities lack effective mechanisms for representing and navigating the model space semantically, particularly when model documentations or names are incomplete or unknown to users, leaving many models effectively undiscoverable~\cite{liang2024whats}.

\smallskip 
\noindent
\introparagraph{Documenting Models} Models that are deployed are usually accompanied with documentation known as model cards~\cite{modelcardpaper}. Model cards contain (among other categories) information on model details, intended use, metrics, training data, and quantitative analyses.
Similar to datasheets~\cite{gebru2021datasheets}, model cards are designed to guide developers in documenting models in a structured way.  Model cards can be and should be augmented with information more similar to nutritional labels~\cite{DBLP:journals/debu/StoyanovichH19} that also include information about fairness and bias in the data (models).  They can be further enriched with lineage and security related documentation~\cite{golpayegani2024ai} such as adversarial attack and related defense measure policies, which are outlined in FactSheets for AI services~\cite{arnold2019factsheets}.In addition to research on enhancing the completeness of model documentation, there remains a critical gap in the verification of  model cards. There is a danger that people could intentionally misinform model users with malicious intent~\cite{poisonGPT}. The state-of-the-art in verifying the documentation of a model is notably in its infancy~\cite{modeldevcheatsheet}. Mithril AICert~\cite{aicert} is developing a certification initiative that verifies whether a model was trained using the specified algorithms and data sets. However, this initiative has some limitations. First, as it is still in development, it is not yet available for production use. Second, the certification process depends on the voluntary participation of the model creators.  Furthermore, the AICert website~\cite{aicertissue} highlights several critical limitations, including the inability to audit training code or data for risks such as backdoors or data poisoning.

\smallskip 
\noindent
\introparagraph{Model Search and Discovery} In ~\Cref{ex:model_search_example}, we present a potential scenario for the Model Search Problem. The current solution pipeline involves a user searching for a relevant model by naming specific models or by typing related keywords like {\tt legal} to find models that either have that word in their name or in their model card. In other words, the search relies on the model's name and documentation. Hence, any sorting of the answer by relevance is just the relevance (prevalence) of the keywords and is not a semantic notion based on the model itself.
Of course, this search may fail if the documentation is incomplete or inaccurate. 

Within open data and enterprise data, researchers have learned that they cannot rely on metadata for datasets to be accurate, complete, or consistent within a data lake~\cite{DBLP:journals/pvldb/NargesianZMPA19}.  As a result, there is a great deal of work on semantic or content-based search in data lakes including join search~\cite{DBLP:journals/pvldb/ZhuNPM16,zhu2019josie,khatiwada2022integrating,2023_dong_deepjoin}, union search~\cite{nargesian2018table,2023_khatiwada_santos,DBLP:journals/pvldb/FanWLZM23,2023_hu_autotus}, and related dataset search~\cite{DBLP:conf/icde/FernandezAKYMS18,DBLP:conf/icde/BogatuFP020}. Machine learning models have revolutionized content-based dataset search.  Important and impactful work has shown how we can use machine learning to create meaningful semantic representations of tuples, columns, and full tables to enhance dataset search and other semantic tasks like data integration and alignment. 

But what about \textit{content-based model search}?   
To the best of our knowledge, this is an area within ML that is in its infancy.   Recent techniques for image models leverage meta learning~\cite{Lu_2023, autoMRM}. 
HuggingGPT~\cite{shen2024hugginggpt} uses an LLM (in their case, ChatGPT) as a controller to decide which models to use based on a user's prompt. This metadata-based search differs from content-based approaches, as the LLM parses the user's prompt into tasks and uses model descriptions to select relevant models. While it allows queries across any modality or domain, it is limited by the LLM's capabilities and the quality of model documentation. Additionally, it may fail when object-centric tags do not fully capture the model. Our vision emphasizes that content-based model search must cover all models in model lakes, including large language models, while ensuring usability through speed and accuracy. 

Another important problem is \textit{related model search}. 
One approach to addressing related model search was explored by~\citet{Lu_2023}, who search for image generative models by using the behavior of another image model as a query. We propose extending such {\em model as query} searches to the model lake to help users identify related models when exploring the model in question. 

\smallskip 
\noindent
\introparagraph{Attribution} An important line of inquiry considers questions related to provenance or attribution~\cite{mei2022provenancemanagement, mu2023model}. These issues pose similar questions to those studied in the database community~\cite{DBLP:journals/sigmod/BunemanT18}:  from where did a generated fact derive or why was a predication made? Building on the concept of data provenance~\cite{buneman2001and}, we can extend the notions of why-, where-, and how-provenance to models, under the name of {\em model attribution}. 
Similar research questions are also posed in model interpretability as part of local and global explanations. As in data provenance, the main issue is in whittling down all the provenance associated with some process (such as a full model) into simple, but useful explanations~\cite{DBLP:journals/sigmod/BunemanT18}. More than two decades of research in the data management community have produced elegant and simple, yet powerful, models for data provenance when the process is a query~\cite{Buneman2008Expressiveness, DBLP:journals/cacm/BunemanDF16} or workflow~\cite{moreau2013prov, DBLP:series/synthesis/2013Moreau}. In large model attribution, the goal cannot be to understand and track all inputs (data, hyperparameters, code, training regimes, ...) that were used to create a specific model output. Hence, the challenge is to find meaningful sets of concepts that can be tracked efficiently and that provide important insights into model behavior.

The training data attribution problem is nontrivial because every aspect of training has the potential to impact every decision of a model. A variety of methods have been developed to estimate influence of training data on model behavior~\cite{koh2017understanding,grosse2023studying,wang2023evaluating}. The current approaches require \textit{extensive use of training data as well as costly analysis of model intrinsics}.  Another approach to the problem is to apply techniques such as membership inference analyses, or membership inference attacks~\cite{shi2024detecting} that ask the question of whether a specific training data item $d$ is present in the training data $\mathcal{D}$, or training data reconstruction methods that extract sets of items from the training data ~\cite{carlini2021extracting, schwarzschild2024rethinkingllmmemorizationlens}.

Insight about the attribution of model behavior can also be studied through \emph{model interpretability} methods: \textit{extraction of relevant knowledge from a model concerning relationships either contained in data or learned by a model}~\cite{definitionInterp}. 
Several works have surveyed techniques and research questions in this area~\citep{DBLP:journals/corr/abs-2408-01416, singh2024rethinking,DBLP:journals/corr/abs-2401-12874, DBLP:journals/corr/abs-2402-10688, DBLP:journals/corr/abs-2207-13243, doshi2017towards, DBLP:conf/dsaa/GilpinBYBSK18}.  These approaches  can be broadly categorized into the following areas.  Local model explanations explain the sensitivity of individual output predictions to local changes in inputs using gradients~\cite{sundararajan2017axiomatic}, masks~\cite{petsiuk2018rise,fong2019understanding}, local models~\cite{ribeiro2016should}, or Shapley values~\cite{ DBLP:journals/corr/LundbergL17, chen2023algorithms}. Global explanations explain the mechanisms of the overall model at the level of attention patterns~\cite{clark-etal-2019-bert}, representations~\cite{belinkov2022probing, zou2023representation}, circuits \cite{elhage2021mathematical,wang2022interpretability}, or neurons~\cite{bau2018identifying,bau2017network,DBLP:conf/emnlp/GevaBFG23}. 
Models can also be designed to be inherently interpretable~\citep{rudin2022interpretable}.
Lastly, datasets can be explained using natural language explanations~\cite{dataWrangleNarayan2022}, data visualizations~\cite{dibia-2023-lida}, or by training inherently interpretable models~\cite{ustun2016supersparse, singh2023augmenting}.

\smallskip 
\noindent
\introparagraph{Model Versions}
Models are valuable assets that can be adapted and reused to create new versions. The original or base version is typically a foundation model --- a pre-trained model that learns general features from its training dataset, denoted as $\mathcal{D}$. By making further adjustments to the training algorithm ($\mathcal{A}$), architecture ($f_*$), or dataset ($\mathcal{D}$), subsequent versions of the base model can be developed. Common $\mathcal{A}$-based modifications include fine-tuning, parameter-efficient tuning, preference tuning, model stitching, and model editing. \textit{Model stitching}, for example, involves altering $f_*$ by combining the architectures of two or more models to create a hybrid model~\cite{lenc2015understanding}. \textit{Model editing}~\cite{gandikota2024unified, DBLP:conf/iccv/OrgadKB23, meng2022memit, meng2022locating, mitchell2022fast, Sinitsin2020Editable} focuses on updating certain facts (e.g., changing the name of the current President of a country), making localized adjustments without retraining the entire model. \textit{Fine-tuning} involves further training a model's parameters,$\theta$, to improve performance on specific task(s) or domain(s). For instance, T5~\cite{raffel2020exploring} is a pre-trained model that has been trained on a collection of large collection of text (about 750 GB) to perform well on a diverse set of tasks. This has been further fine-tuned with the MIMIC-III~\cite{johnson2016mimic} and IV~\cite{johnson2023mimic} dataset to form Clinical-T5~\cite{lu-etal-2022-clinicalt5} to perform better for medical domain-related tasks. 
In addition to the traditional fine-tuning strategy, \textit{parameter-efficient fine-tuning} methods have emerged to reduce computational overhead by freezing most of the model's parameters, only updating a small subset necessary for fine-tuning. For instance, Low Rank Adaption (LoRA)~\cite{hu2022lora} is a parameter-efficient fine-tuning method to adapt various models by only updating a low-rank subset~\cite{hu2022lora}. \textit{Preference tuning} is another advanced technique, as seen in ChatGPT~\cite{openaichat} and InstructGPT~\cite{ouyang2022traininglanguagemodelsfollow}, which integrates human feedback into the fine-tuning process~\cite{winata2024preferencetuninghumanfeedback}. Lastly, due to the emergent ability of models to perform tasks without training~\cite{dong2022survey, 10.1145/3560815}, newer models leverage \textit{prompting} as a way to control content generation without needing further updates to their parameters.

Information about model versioning can be inferred even if the model history is unavailable.  For example, Mu et al.~\cite{mu2023model} propose a data- and model-driven method to encode ``Model DNA" for identifying if a model is a pre-trained version of another, assuming both share the same architecture and training data. However, more challenging cases arise when the target model has a different architecture or is trained on only a subset of the source model's parameters. Hugging Face recently introduced new metadata fields in their model cards, enabling users to specify the base model and explain how it has been modified. This metadata generates a model tree, linking related models by their extensions. However, its accuracy depends on reliable documentation, and older models lack this data. Research on reconstructing and verifying relationship is emerging~\cite{horwitz2024origin,sun2025idiosyncrasieslargelanguagemodels,zhu2025independencetestslanguagemodels}, such as ~\citet{horwitz2024origin}'s approach using weight similarities, though this approach is limited to known models with a single base version.

\smallskip 
\noindent
\introparagraph{Privacy and Safety} Models are vulnerable to the disclosure of private information~\cite{7958568,carlini2021extracting} and adversarial results~\cite{zou2023universal,li-etal-2025-revisiting} when attacked. Hence, initiatives such as Privacy Preserving Machine Learning~\cite{ruehle2021privacy} exist to understand, measure, and mitigate such risks. As a result, techniques such as differential privacy~\cite{dwork2006calibrating, yan2024protecting}, data sanitization~\cite{staab2025language, dou-etal-2024-reducing} and robust prompt optimization~\cite{zhou2024robust} have been utilized to defend against such attacks. These methods generally aim to obscure or eliminate private information while detecting and preventing attempts to jailbreak or manipulate the model and its output. However, this can create a false sense of privacy as defense techniques can continue to be compromised with other attack schemes~\cite{xin2024a}. 

Beyond privacy, ensuring AI systems are fundamentally safe is another critical challenge. Community-driven efforts focus on building safe AI~\cite{ortega2018building}, aiming to specify, verify, and ensure that the models behave as intended. A recent approach leverages neuroscience concepts, such as representation engineering, to enhance AI transparency~\cite{zou2023transparency, mineault2024neuroaiaisafety} and improve our understanding of traits like honesty, power seeking tendencies, and morality in models. However, this direction is still in its early stages, with open challenges including determining the range of representations of concepts and functions that can be extracted from models and developing scalable evaluation methods to better inform users about model safety.

\smallskip 
\noindent
\introparagraph{Benchmarking}
Benchmarking plays a crucial role in evaluating model performance for specific tasks and remains a well-established research area, however benchmarks for newer topics such as model attribution and versioning are urgently needed. Developers frequently report model performance using standardized benchmarks, making it a routine part of model assessment. Broadly, there are two primary types of benchmarking: (1) evaluating how well a model, $\mathcal{M}$, approximates a ground truth distribution, and (2) assessing a model’s performance against a targeted evaluation metric. In classification tasks, models are typically evaluated using accuracy, as ground-truth labels provide a direct assessment of the model's output. Additionally, confusion matrices offer insights into the types of errors made. For text generation tasks, perplexity is a widely used metric, and several popular benchmarks exist to assess performance~\cite{mmlu, hendrycks2021measuring, rein2024gpqa}. 
In the case of image generation, the Fréchet Inception Distance (FID)~\cite{heusel2017gans} is a common metric and COCO~\cite{lin2014microsoft} and VQAv2~\cite{Goyal_2017_CVPR} are some notable benchmarks in this domain.
Beyond traditional performance metrics, benchmarking also considers biases related to protected attributes~\cite{schramowski2023safe}, as well as the environmental impact~\cite{lacoste2019quantifying} of model training and deployment. Model lake benchmarks lack large-scale, publicly available datasets that mimic realistic conditions of diverse models in model lakes. There has been preliminary efforts by ~\citet{Lu_2023}, where they released a benchmark for model search with 259 publicly available image generative models and 1000 customized text-to-image diffusion models. Model Zoo~\cite{schurholt2022model} is another relatively large-scale dataset, but it is limited to vision models. Creating large-scale model lake benchmark  that integrates various modalities (such as text, images, and audio) remains a research challenge.
\section{Research Roadmap}\label{sec:roadmap}

\begin{figure*}
    \centering
\includegraphics[width=\textwidth]{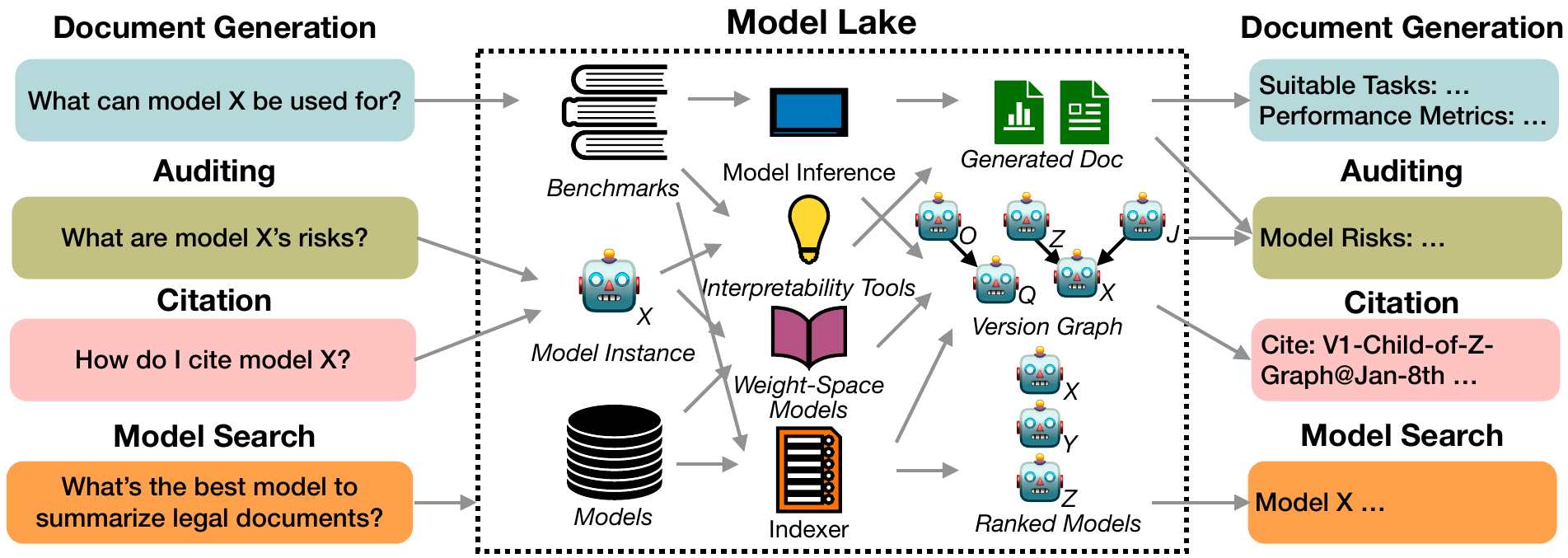}
    \caption{Model Lakes Design. A model lake  stores models and processes them using techniques, like inference, interpretability, weight-space modeling and indexing to support various user interactions. It generates outputs like version graphs, model cards and ranked models, refining them into human-readable results, as shown on the figure's right side.}
    \label{fig:model-lakes-framework}
\end{figure*}

Building on current research and identified gaps outlined in ~\Cref{sec:sota}, this section presents a vision to address these challenges. A model lake is illustrated in ~\Cref{fig:model-lakes-framework}, where rather than APIs, we envision data scientists interaction with the lake using queries.

\smallskip
\noindent
\introparagraph{Benchmarking}
As discussed in ~\Cref{sec:sota}, benchmarking, while a well-established research area, is under-explored in model lakes. An important topic is the development of lifelong benchmarks~\cite{prabhu2024lifelongbenchmarksefficientmodel} that can address increasingly complex and novel scenarios as models continue to evolve in capability and diversity. In addition, there is a need to develop benchmarks specific to model lake tasks. For instance, to advance research in model attribution, a comprehensive \textit{benchmark dataset} is needed—one that includes labeled model parameters, architectures, and detailed transformation records (e.g., fine-tuning, model editing). This benchmark can be extended to model versioning by adding data that tracks the previous and subsequent versions of models.

\smallskip
\noindent
\introparagraph{Model Inference}
Deploying effective solutions for model lake tasks is a challenge with respect to \textit{usability} and \textit{scalability}. To improve usability, the model inference component involves \textit{identifying appropriate benchmarks and generating relevant prompts, as well as selecting suitable models (target models or meta-models)} to address a user query. While users can manually run prompts and select models, this approach is prone to errors and suboptimal outcomes, especially if users lack the expertise to use models effectively. For example, a classifier's behavior may be misinterpreted if a user does not understand the type of data is was trained or the input it expects. To mitigate this, the proposed model lake tasks can incorporate additional perspectives, such as intrinsic model properties (e.g., weights), to provide insights and guide users towards a more accurate interpretations and applications. This search and generation process can also be automated using an AI agent. By applying benchmarks and model(s) in question, we combine research efforts of benchmarking and attribution questions, which can explain a model's behavior and output.

\smallskip
\noindent
\introparagraph{Indexer}
A central component of a model lake is the indexer, which would be used to embed and provide scalable sublinear search over the model embeddings. The indexer can use different ranking functions tailored to the task.
\citet{DBLP:journals/corr/abs-2406-07536, 9008292} have introduced an approach for generating model and task embeddings. 
However, search methods must scale to handle millions of models and adapt to newer models with advanced capabilities~\cite{wei2022emergent}.  Indices like HNSW (Hierarchical New Small World)~\cite{DBLP:journals/pami/MalkovY20}, have proven effective in practice in indexing high-dimensional embeddings enabling fast nearest-neighbor search (including over data lakes~\cite{DBLP:journals/pvldb/FanWLZM23}.  However, HNSW provides no formal guarantees on correctness and its use in model lakes remains under-explored. Effective embedding of models is crucial for accurate comparison and ranking by the indexer. ~\citet{DBLP:journals/corr/abs-2406-07536} explores this but there work is not inclusive of all model types. A robust system should support diverse embeddings to ensure indexing effectiveness. Many of the model lake tasks will benefit from  hybrid approach, that indexes both metadata and model embeddings -- for example, related model search can combine well-chosen model embeddings representing important intrinsic model features with search over verified models cards. Similarly, in versioning, embeddings and their associated rankings can aid in identifying parent-child relationships and assessing the distinctiveness of each model relative to others.

\smallskip
\noindent
\introparagraph{Weight-Space Modeling} Weight-Space modeling is a hyper-representation learning approach where a neural network is trained to process weights of other models~\cite{eilertsen2020classifying}. This method can be useful for making distinctions between models, especially in complex scenarios like model stitching, where similar models with multiple shared parent models need to be distinguished. This is a promising direction.   ~\citet{10.5555/3692070.3694629}, for instance, reveals a linear connection between fine-tuned models. This approach could also facilitate dynamic selection of benchmarks for performance measurement by learning from previous relationships between datasets and models. The primary challenge in weight-space modeling is identifying the most relevant aspects of the model for training another model to uncover patterns, while simultaneously considering the weight-space model architecture~\cite{pmlr-v202-navon23a}. It is also crucial to ensure sufficient data diversity to avoid overfitting or underfitting.

\smallskip
\noindent
\introparagraph{Interpretability} Interpretability methods can be most useful for the model attribution task where users must understand the origin of model behavior, for example, when detecting knowledge changes~\cite{youssef2024detectingeditedknowledgelanguage}, or when unlearning learned knowledge~\cite{jang-etal-2023-knowledge, eldan2023whos,gandikota2023erasing, belrose2023leace, kumari2023conceptablation}. Attribution questions also apply to the source model's architecture. Approaches like circuit discovery~\cite{wang2022interpretability, conmy2023automated, elhage2021mathematical} can help identify the computational origin of model behavior.
\textit{Model inversion} can be used to recover an input prompt given an output~\cite{morris-etal-2023-text, morris2024language, huang2024inversionviewgeneralpurposemethodreading}. These methods are also a promising route for understanding the impact of training data on internal model states. In recent work, ~\citet{sharkey2025openproblemsmechanisticinterpretability} present a list of open questions in interpretability, many of which are relevant to the problem of model lakes.

\smallskip
\noindent
\introparagraph{Holistic Management of Models and Data}
Effective model lakes need to integrate advances in data lakes in a holistic way given the reliance of models on data as their fuel.
Many model management tasks include (in part) an analysis of training or input data. Hence, traditional data lake concerns such as data provenance, data versioning, and related issues still apply. Thus, in addition to new tasks specific to model lakes, we must also account for data lake tasks when dealing with the data used by or generated from these models. And integrating these tasks will be important.  As a simple example, when searching for models trained on a dataset, users may want to find models trained on versions of the dataset.

\section{MODEL LAKE TASKS: APPLICATIONS}\label{sec:applications}
\smallskip 
\noindent
\introparagraph{Documentation Generation}
To streamline the creation of detailed model cards, engineers can leverage model lake tasks to generate a rough draft of the required documentation. This application is similar to discovering and annotating metadata in data lakes~\cite{croissant, sawadogo2021data, 10.1145/3472163.3472273, DBLP:conf/vldb/KoriniB23, FanTutorial2023}. Here's how the process might work: upon uploading a model to the model lake, state-of-the-art techniques for tasks like attribution, versioning, benchmarking, and others can automatically analyze and map the model's relationships to other models in the model lakes. The engineer can review and either accept or modify these generated mappings, especially if any information appears inaccurate, creating an initial version of the model details section. Additionally, the engineer can assess the model's robustness by testing its performance against relevant benchmarks, including metrics like task accuracy, environmental impact, and more. Based on these test results, key sections of the model card, such as intended use and performance metrics, can be auto-populated for a faster, more reliable drafting process. It can alert developers to model risks, as shown by~\citet{wang2024mitigatingdownstreammodelrisks}, who demonstrate how model versioning helps warn downstream model users of upstream model risks.

\smallskip 
\noindent
\introparagraph{Auditing}
Policymakers have recently proposed AI safety and compliance regulations~\cite{house2022, 925786, euaiact} aimed at fostering more responsible and accountable AI models~\cite{casper2024blackbox}. The model document generation application procedure can be repurposed for auditing by creating a template questionnaire~\cite{golpayegani2024ai} and using the information from the model lake to generate a draft response with proof or explanation about how a requirement is fulfilled. This process can incorporate privacy-related technical solutions, where insights from model lake tasks help identify vulnerabilities across related models and their successor versions. It can also aid in attributing sensitive data that the model may have access to, highlighting potential exposure risks.

\smallskip 
\noindent
\introparagraph{Data and Model Citation} \textit{Data citation} helps stakeholders identify the source, ownership, and authorship of the data used for a particular analysis. It is important to use proper data citation because the structure and contents of the database can evolve~\cite{DBLP:journals/cacm/BunemanDF16}. Hence, this problem is also extended to machine learning, since a large part of model training is its dataset. Thus, the task of citing data for data lakes remains relevant for documenting the training data used in the creation of the model. Similarly, \textit{model citation} is essential, as users can further train the model or use its outputs for consumption or additional training. One proposed solution to identify generated output is the use of watermarks~\cite{pmlr-v202-kirchenbauer23a}. In addition, \textit{model versioning} tasks provide crucial documentation, allowing researchers, engineers, and developers to refer to the exact version of the model used for training or content generation. If a particular model is used, the platform would refer to its versioning graph and generate a citation with the model version and timestamp of the graph. Upon any updates of the graph, a new citation would be generated with the updated version and timestamp. This would be useful for also citing any generated content from this model.

\smallskip 
\noindent
\introparagraph{Model Search} In ~\Cref{ex:model_search_example}, we illustrate a potential scenario where a user seeks the most suitable model for summarizing and simplifying legal documents. As the model search task within the model lake evolves, we aim for users to be able to write declarative queries and retrieve a set of models ranked by their suitability for the specified task.  Query example include "Find all models trained on this corpus of US Supreme Court cases"  or "Find models that out perform Model X on Benchmark Y". Given a search task, the model lake framework can map the task function to a suitable indexer and run that indexer to retrieve top-ranked models. Whether the user is a technical or non-technical consumer, researcher, or engineer, they would be able to click on a model to view its model card. Attribution reveals how training data or learned concepts influence outputs, while version management clarifies the model's training process, lineage, and differences, enhancing transparency. Benchmarking evaluates the model's robustness on related tasks, addressing completeness. Together, these allow users to make informed model choices.
\section{Conclusion}
The database community has responded to the ``Big Model" revolution by proposing platforms like Agora~\cite{Agora20}, that manage data-related assets, including models, datasets, software, and compute resources in a coherent ecosystem. But the model-specific support in such an ecosystem must be expanded to include general methods for managing and finding models, support that can make such systems fully functional model lakes. It will be important that the methods generalize, irrespective of how many models we are trying to understand, what architectures the models use, how they are trained, or how we wish to query or search the models. We call on the database community to contribute to the vision of model lakes, supporting users to more easily find relevant models and to better understand those models. Our vision is for a fundamentally new platform that extends and integrates work on data/model attribution, data/model search, and data/model version management.
\begin{acks}
KP and RM were supported in part by NSF award numbers IIS-2107248,  IIS-1956096, IIS-2325632, and KP and DB by a grant from Open Philanthropy.
\end{acks}
\clearpage
\bibliographystyle{ACM-Reference-Format}
\bibliography{main}

%
% ...

\end{document}